\documentclass[submission, Proceedings]{SciPost}

\hypersetup{
    colorlinks,
    linkcolor={red!50!black},
    citecolor={blue!50!black},
    urlcolor={blue!80!black}
}

\usepackage[bitstream-charter]{mathdesign}
\DeclareSymbolFont{usualmathcal}{OMS}{cmsy}{m}{n}
\DeclareSymbolFontAlphabet{\mathcal}{usualmathcal}

\begin{document}

\begin{center}{\Large \textbf{
Shadow generalized parton distributions: a practical approach to the deconvolution problem of DVCS
}}\end{center}

\begin{center}
V. Bertone\textsuperscript{1},
H. Dutrieux\textsuperscript{1$\star$},
C. Mezrag\textsuperscript{1},
H. Moutarde\textsuperscript{1} and
P. Sznajder\textsuperscript{2}
\end{center}

\begin{center}
{\bf 1} IRFU, CEA, Université Paris-Saclay, F-91191 Gif-sur-Yvette, France
\\
{\bf 2} National Centre for Nuclear Research (NCBJ), Pasteura 7, 02-093 Warsaw, Poland
\\
* herve.dutrieux@cea.fr
\end{center}

\begin{center}
\today
\end{center}


\definecolor{palegray}{gray}{0.95}
\begin{center}
\colorbox{palegray}{
  \begin{tabular}{rr}
  \begin{minipage}{0.1\textwidth}
    \includegraphics[width=22mm]{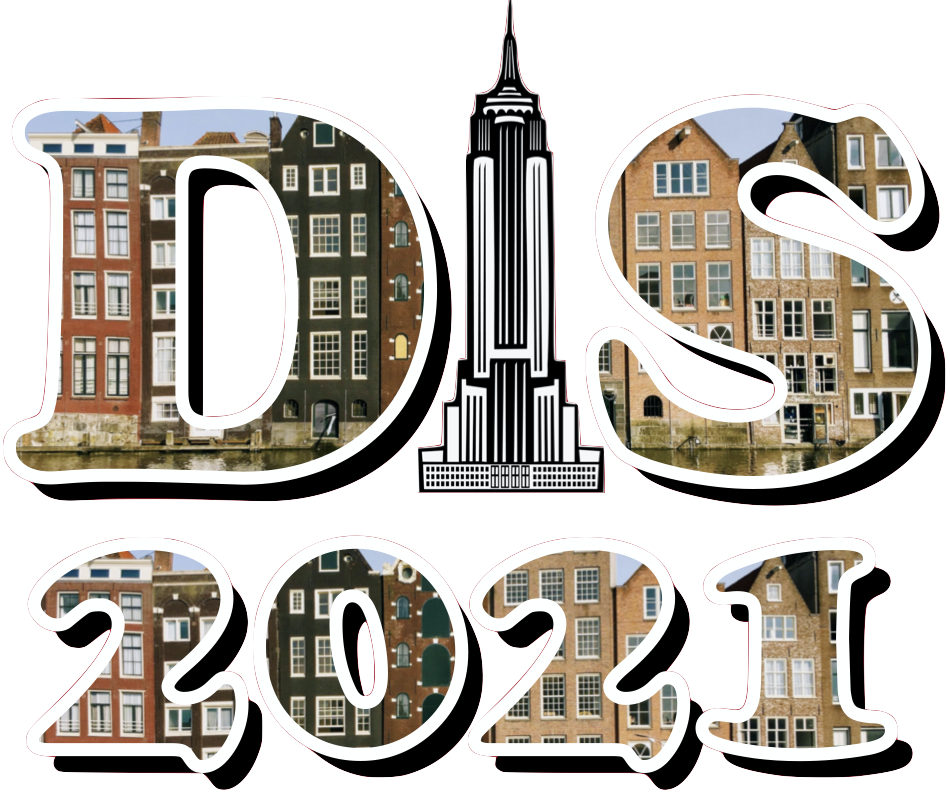}
  \end{minipage}
  &
  \begin{minipage}{0.75\textwidth}
    \begin{center}
    {\it Proceedings for the XXVIII International Workshop\\ on Deep-Inelastic Scattering and
Related Subjects,}\\
    {\it Stony Brook University, New York, USA, 12-16 April 2021} \\
    \doi{10.21468/SciPostPhysProc.?}\\
    \end{center}
  \end{minipage}
\end{tabular}
}
\end{center}

\section*{Abstract}
{\bf
Deeply virtual Compton scattering (DVCS) attracts a lot of interest due to its sensitivity to generalized parton distributions (GPDs) which provide a rich access to the partonic structure of hadrons. However, the practical extraction of GPDs for this channel requires a deconvolution procedure, whose feasibility has been disputed. We provide a practical approach to this problem based on a next-to-leading order analysis and a careful study of evolution effects, by exhibiting \textit{shadow} GPDs with arbitrarily small imprints on DVCS observables at current and future experimental facilities. This shows that DVCS alone will not allow for a model independent extraction of GPDs and a multi-channel analysis is required for this purpose.
}


\section{Introduction}
\label{sec:intro}

Generalized parton distributions (GPDs), introduced in Refs. \cite{Mueller:1998fv, Ji:1996ek, Ji:1996nm, Radyushkin:1996ru, Radyushkin:1997ki}, are involved in the description of a wide class of exclusive reactions thanks to factorization theorems \cite{Radyushkin:1997ki, Collins:1998be, Ji:1998xh}. Among those processes, deeply virtual Compton scattering (DVCS) is considered as a particularly promising channel thanks to its clean theoretical interpretation and the wealth of data already available for fits \cite{Moutarde:2018kwr, Moutarde:2019tqa, Cuic:2020iwt}. DVCS observables are parametrized in terms of Compton form factors (CFFs) which write in the framework of collinear factorization as the convolution of a coefficient function, which can be computed in perturbation theory, and a non-perturbative GPD. Deconvoluting this relation is necessary to extract GPDs from experimental results in a model independent way. It has been argued that evolution would play a crucial role in the deconvolution procedure \cite{Freund:1999xf}, but no full-fledged theoretical argument beyond the leading order (LO) component of the coefficient function, or phenomenological feasibility proof has been put forward. 

In this note, we first remind the expression of the convolution of interest and a key property of GPDs useful for our study. Then we detail how the concept of shadow GPDs introduced in Ref.~\cite{Bertone:2021yyz} can be used to determine the feasibility of a deconvolution procedure and present the numerical results of evolution of next-to-leading order (NLO) shadow GPDs. Finally we discuss their consequences for GPDs extraction.

\section{The DVCS convolution} 

For the sake of simplicity, the GPD $H(x, \xi, t, \mu^2)$ and its associated CFF $\mathcal{H}(\xi, t, Q^2)$ are considered in the following, but the discussion can be easily extended to other GPDs $E$, $\widetilde{H}$ or $\widetilde{E}$. Since $t$ acts as a mere parameter in this discussion, it is omitted. The CFF $\mathcal{H}$ can be written as the sum of scheme-dependent quark $\mathcal{H}^q$ and gluon $\mathcal{H}^g$ contributions. Assuming that $\mathcal{H}^q$ is known, we are left with the following relation to deconvolute
\begin{equation}
    \mathcal{H}^q(\xi, Q^2) = \int_{-1}^{1} \frac{\mathrm{d}x}{2\xi} T^q\left(\frac{x}{\xi}, \frac{Q^2}{\mu^2}, \alpha_S(\mu^2)\right)H^{q(+)}(x,\xi,\mu^2) \,,
    \label{eq:convol}
\end{equation}
where $H^{q(+)}(x,\xi,\mu^2) = H^q(x,\xi,\mu^2) - H^q(-x,\xi,\mu^2)$ is the singlet component of the GPD and $T^q$ the coefficient function which can be computed in perturbation theory. The expression of $T^q$ up to NLO can be found \textit{e.g.} in Refs.~\cite{Belitsky:1999sg, Pire:2011st}. Since the variable $x$ is integrated out in Eq.~\eqref{eq:convol}, one could intuitively expect that the information contained in $\mathcal{H}^q(\xi, Q^2)$ is insufficient to fully recover $H^{q(+)}(x,\xi,\mu^2)$. However, the $\mu^2$ dependence of the GPD is constrained by evolution equations, and the $x$ and $\xi$ dependences are tied by the requirement that $x$ Mellin moments of the GPD are polynomials in $\xi$ \cite{Ji:1998pc,Radyushkin:1998bz}, \textit{i.e.}
\begin{equation}
    \int_{-1}^{1} \mathrm{d}x\,x^n H^{q(+)}(x,\xi,\mu^2) = \sum_{\substack{k = 0 \\ \mathrm{even}}}^{n+1} H^{q,n}_k(\mu^2)\, \xi^k \,.
    \label{eq:pol}
\end{equation}
A convenient representation equivalent \cite{Chouika:2017dhe, Chouika:2017rzs} to Eq.~\eqref{eq:pol} is obtained by writing $H^{q(+)}$ as the Radon transform of the sum of a double distribution (DD) $F^{q(+)}$ and a function $D^q$ called the D-term
\begin{align}
    H^{q(+)}(x,\xi,\mu^2) = \int \mathrm{d}\Omega [F^{q(+)}(\beta, \alpha, \mu^2) + \xi \delta(\beta)\,D^q(\alpha,\mu^2)]\,,
    \label{eq:radon}
\end{align}
with $\mathrm{d}\Omega =\mathrm{d}\beta\,\mathrm{d}\alpha\,\delta(x-\beta-\alpha\xi)$, where $|\alpha|+|\beta|\leq 1$.

\section{Shadow generalized parton distributions}

We first focus on a given scale $\mu_0^2$. To address the question of deconvoluting Eq.~\eqref{eq:convol} at that scale while respecting the important property of Eq.~\eqref{eq:pol}, we exhibit explicit non-vanishing DDs $F^{q(+)}(\beta, \alpha, \mu_0^2)$, coined shadow DDs, so that the convolution of Eq.~\eqref{eq:convol} is exactly zero, doing so for the first time for the NLO expansion of the coefficient function $T^q$. Furthermore, since the forward limit $\xi \rightarrow 0$ of the GPD $H$ gives back a usual parton distribution function (PDF) which is experimentally well known, we require that our shadow DDs give rise via Eq.~\eqref{eq:radon} to shadow GPDs with also a vanishing forward limit. The existence of shadow DDs shows that the problem of extracting GPDs from DVCS and PDF data strictly admits an infinite number of solutions at a given scale. Indeed, any shadow GPD can be freely added to a phenomenologically relevant GPD without altering its NLO CFF and forward limit.

The demonstration of the existence of such shadow DDs, detailed in Ref.~\cite{Bertone:2021yyz}, makes use of the remarkable properties of CFFs computed from DDs which write as simple polynomials of a given order $N$ in $\beta$ and $\alpha$. For such DDs, successively the LO component of the CFF, the NLO collinear component and the one-loop component can be decomposed into a finite partial fraction expansion at $\xi = \pm 1$ of $\sim N$ terms and an essential singularity which is systematically cancelled by the previous term. As an example, we provide the result for the imaginary part of the one-loop component, which can be written as
\begin{equation}
\textrm{Im}\,T^q_{one-loop} \otimes H^{q(+)} = \log\left(\frac{1-\xi}{2\xi}\right) \textrm{Im}\,T^q_{coll} \otimes H^{q(+)} + \frac{e_q^2C_F}{4} \sum_{l=1}^{N-1} \frac{k_l(H^{q(+)})}{(1+\xi)^l}\,,
\label{eq:oneloop}
\end{equation}
where we have noted the one-loop component to the CFF as $T^q_{one-loop} \otimes H^{q(+)}$ to remind that it is the convolution of the GPD with the one-loop component of the coefficient function $T^q$, and done the same for the NLO collinear component of the CFF $T^q_{coll} \otimes H^{q(+)}$. The $k_l$ are a set of linear functions of the coefficients characterizing the polynomial DD which gives rise to $H^{q(+)}$ via Eq.~\eqref{eq:radon}. Therefore, the last term of Eq.~\eqref{eq:oneloop} can be cancelled by finding the kernel of the linear system formed by the $N-1$ linear functions of the polynomial coefficients. Since a polynomial DD of order $N$ in $\beta$ and $\alpha$ has $\sim N^2$ such coefficients, it is clear this system admits a kernel of arbitrary large dimension as $N$ increases. As for the first term of Eq.~\eqref{eq:oneloop}, it has itself been cancelled in the previous step of the demonstration by similar means. Iteratively, we are able to cancel each term of the CFF thanks to their properties when using polynomial DDs. The forward limit is also cancelled similarly. 

\begin{figure}[h]
\centering
\includegraphics[width=0.5\textwidth]{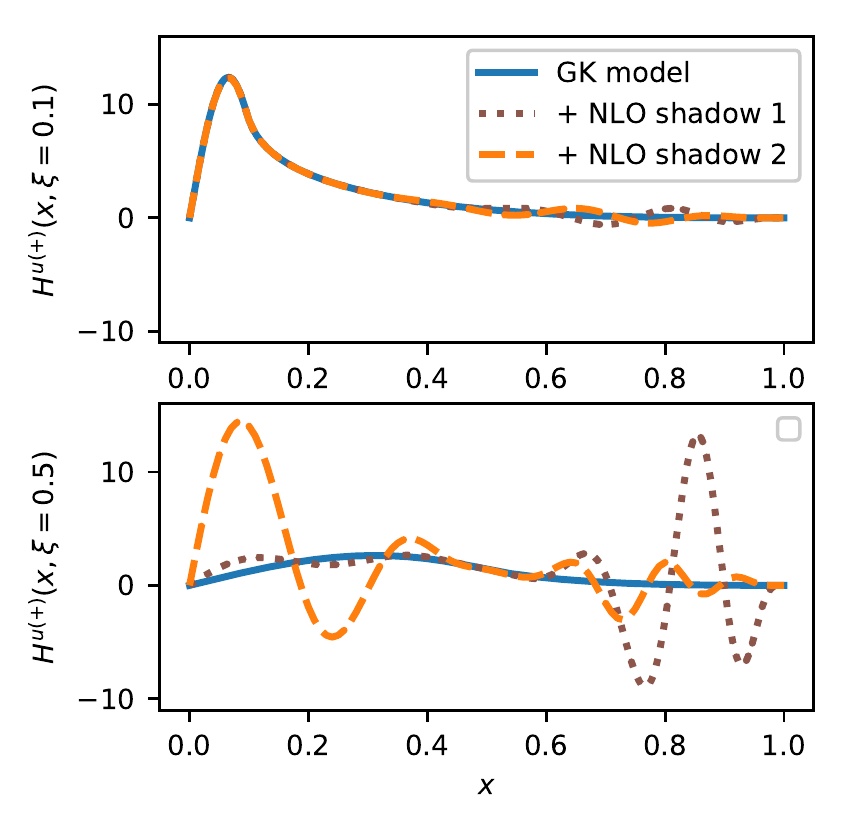}
\caption{$H^{u(+)}$ as a function of $x$ for $\xi = 0.1$ and 0.5, $t = -0.1$ GeV$^2$ and $\mu_0^2 = 1$ GeV$^2$. Solid blue: Goloskokov-Kroll (GK) model. Dashed orange and dotted brown: GK model with the addition of two different NLO shadow GPDs. In all cases one obtains exactly the same NLO CFF and forward limit at scale $\mu_0^2$.}
\label{fig:GK}
\end{figure}

Examples of shadow GPDs found by this procedure are given in Fig.~\ref{fig:GK}, where they have been added to the Goloskokov-Kroll model \cite{Goloskokov:2005sd, Goloskokov:2007nt, Goloskokov:2009ia}.

\begin{figure}[h]
\centering
\includegraphics[width=0.5\textwidth]{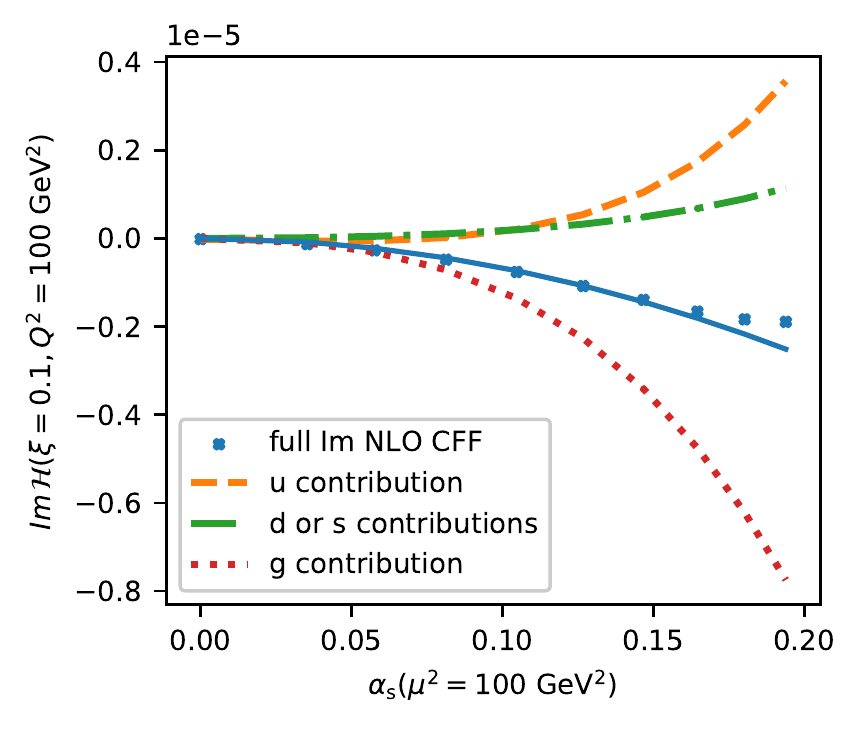}
\caption{Imaginary part of the NLO CFF $\mathcal{H}(\xi = 0.1, Q^2 = 100~\mathrm{GeV}^2)$ evaluated with the NLO shadow GPD 1 shown in Fig.~\ref{fig:GK} and $H^{d(+)} = H^{s(+)} = H^{g(+)} = 0$ at $\mu_0^2 = 1$ GeV$^2$. The blue dots correspond to computations made with different values of $\alpha_S(\mu^2 = 100~\mathrm{GeV}^2)$ and the solid blue line results from a quadratic fit to the first seven points. The dashed orange, dash-dotted green and dotted brown lines indicate $u$, $d$ or $s$, and $g$ contributions to the CFF, respectively.}
\label{fig:evol}
\end{figure}

The effect of evolution on a realistic setting for future collider facilities, like the US electron-ion collider (EIC) \cite{Accardi:2012qut, AbdulKhalek:2021gbh}, Chinese electron-ion collider (EIcC) \cite{Anderle:2021wcy}, and large hadron-electron collider (LHeC) \cite{AbelleiraFernandez:2012cc}, is probed using APFEL++ \cite{Bertone:2013vaa,Bertone:2017gds,Bertone:2021} and PARTONS \cite{Berthou:2015oaw}. Choosing $\xi = 0.1$ and a range in $Q^2$ from 1 to 100 GeV$^2$, we vary the strong coupling $\alpha_S(\mu^2 = 100~\mathrm{GeV}^2)$ from 0 to its typical $\overline{\textrm{MS}}$ value in Fig.~\ref{fig:evol}. As expected, the CFF varies as $\mathcal{O}(\alpha_S^2)$ since we have specifically cancelled its LO and NLO components. The numerical value of the obtained CFF is so small (less than $10^{-5}$) that it will be hidden in typical statistical and systematical uncertainties of current and future experiments, such that even this lever arm in evolution is not enough to meaningfully change our conclusion on the feasibility of the DVCS deconvolution.

\section{Conclusion}

The existence of shadow GPDs with a null forward limit and negligible contributions to CFFs over kinematic domains relevant for current and future experimental facilities is a challenge in several aspects. Since data are extracted at non-zero skewness $\xi$, the extrapolation towards zero skewness required for proton tomography may suffer from residual dependence on shadow GPDs. The role of the D-term has been omitted so far, but the concept of shadow GPD could be extended to GPDs with a vanishing subtraction constant in the dispersion relation formalism, but still a non-vanishing D-term. This extension would be crucial for the extraction of proton mechanical properties, which has been demonstrated to be highly sensitive to experimental uncertainties in existing data \cite{Dutrieux:2021nlz}. The reduction of experimental uncertainty on CFFs themselves is of course a pre-requisite to a lesser model dependent extraction of GPDs, which will greatly benefit from the exploration of new observables \cite{Dutrieux:2021ehx} and kinematic ranges.

The NLO time-like Compton scattering (TCS) observables and LO deeply virtual meson production (DVMP) observables computed from NLO shadow GPDs are equally zero. We foresee that our reasoning can be extended to guarantee the existence of shadow gluon GPDs and more generally of shadow GPDs at any finite order in the perturbative expansion of the DVCS coefficient function. The first few moments of GPDs computed on the lattice will not change this picture either. However, a multi-channel analysis including \textit{e.g.} DVMP at higher order or processes with a richer structure in terms of kinematic variables, like double deeply virtual Compton scattering (DDVCS) for instance, could bring quantitative constraints on shadow GPDs. Lattice computation in the $x$ space \cite{Constantinou:2020pek} and the implementation of more theoretical constraints, like positivity bounds \cite{Pire:1998nw, Diehl:2000xz, Pobylitsa:2001nt, Pobylitsa:2002gw}, also call for subsequent studies.


\paragraph{Funding information}
This project was supported by the European Union's Horizon 2020 research and innovation programme under grant agreement No 824093. This work was supported by the Grant No. 2019/35/D/ST2/00272 of the National Science Center, Poland. The project is co-financed by the Polish National Agency for Academic Exchange and by the COPIN-IN2P3 Agreement.






\bibliography{bibliography.bib}

\nolinenumbers

\end{document}